# Developing a Robust Migration Workflow for Preserving and Curating Hand-held Media

Angela Dappert
Digital Preservation Coalition
Innovation Centre, York University
Science Park, Heslington,
York YO10 5DG

angela@dpconline.org

Andrew Jackson
The British Library
Boston Spa, Wetherby
West Yorkshire, LS23 7BQ, UK
+44 (0) 1937 546602

Andrew.Jackson@bl.uk

Akiko Kimura
The British Library
96 Euston Road
London, NW1 2DB, UK
+44 (0) 20 7412 7214

Akiko.Kimura@bl.uk

## ABSTRACT
Many memory institutions hold large collections of hand-held media, which can comprise hundreds of terabytes of data spread over many thousands of data-carriers. Many of these carriers are at risk of significant physical degradation over time, depending on their composition. Unfortunately, handling them manually is enormously time consuming and so a full and frequent evaluation of their condition is extremely expensive. It is, therefore, important to develop scalable processes for stabilizing them onto backed-up online storage where they can be subject to high-quality digital preservation management. This goes hand in hand with the need to establish efficient, standardized ways of recording metadata and to deal with defective data-carriers. This paper discusses processing approaches, workflows, technical set-up, software solutions and touches on staffing needs for the stabilization process. We have experimented with different disk copying robots, defined our metadata, and addressed storage issues to scale stabilization to the vast quantities of digital objects on hand-held data-carriers that need to be preserved. Working closely with the content curators, we have been able to build a robust data migration workflow and have stabilized over 16 terabytes of data in a scalable and economical manner.

## Categories and Subject Descriptors
H.3.2 [**Information Storage**]; H.3.6 [**Library Automation**]; H.3.7 [**Digital Libraries**]; I.7 [**DOCUMENT AND TEXT PROCESSING**]; J.7 [**COMPUTERS IN OTHER SYSTEMS**]

## 1 INTRODUCTION
Digital objects typically undergo several processing steps before they can be considered well-managed. For practical purposes, we use 4 coarse stages to describe how well managed a digital object or collection is. This is a simplification of the criteria one might use in a full risk assessment. For example, it conflates several criteria and the states don't necessarily develop in exactly this order. Nevertheless, it gives a useful pragmatic classification to assess the preservation quality of a set of collections.

The lowest-rated category is that of handheld data-carriers and is considered absolutely unsatisfactory for digital preservation purposes. Hand-held data-carriers tend to decay quickly (particularly writable or re-writable carriers that use optical dyes) and their rendering devices become obsolete relatively quickly. It is not always feasible to properly quality assess them upon arrival, due to the high manual overhead associated with checking each item. This also means that there is often only a single physical copy held, with no backup in case it should get damaged. Online storage is considered more resilient, easier to check for deterioration, easier to refresh if deterioration is detected and easier to manage remotely. For all these reasons it is very desirable to move digital objects as quickly as possible into the second category, the bit-stable category.

Table 1. 4-category digital object status progression

| Unsatisfactory object status | Bit-stable object status | Content stable object status | Archival object status |
|---|---|---|---|
| Hand-held carriers | Content has been transferred onto managed hard disk storage. Storage is backed up. Checksums have been calculated. | Content has been QA'ed. Metadata has been produced and QA'ed. File formats have been identified. Representation Information has been deposited. | Automatic check for corruption via checksums. Automatic replication over remote locations. Digital signatures. Integration with the catalogue. |
|  | Step 1 | Step 2 | Step 3 |

33

To do this, the digital objects must be transferred onto managed hard disk storage and storage that is backed up. We refer to this migration process as data-carrier stabilization. In addition to preservation concerns there are other functional requirements to hold digital objects on managed hard disk storage: to render the content searchable, to provide remote access, to replicate content in several locations, etc. The next two steps of creating content stable and archival object status, as described in Table 1, are just as important. But this paper will focus on issues involved in this first step of creating bit-stable objects.

The British Library has numerous collections, which exist to a significant degree on hand-held data-carriers. They vary in their characteristics, their preservation needs, and the eventual usage of their content. For example, in the Endangered Archives Programme [27], funded by Arcadia [4], valuable digitized material is sent to the British Library from projects all over the world. Due to local variability in technology, the content and the data-carriers on which the content is transmitted vary greatly; they are not always satisfactorily QA'ed *before* they are sent in, file formats vary, and they don't arrive uniformly catalogued. This means that content and metadata often need to be edited and restructured by curators to create a consistent collection, and the stabilized material needs to be easily findable, accessible by curators, and needs to be associated with metadata that may be held offline. While the data-carriers themselves are not valuable artifacts that need to be preserved, their content is and the distribution of the content over several data-carriers is sometimes associated with important semantic distinctions. In contrast, most of the CDs and DVDs in the Sound & Vision collection are published, mass-replicated disks, and have a much higher life expectancy than (re)writable disks. Furthermore, the disks themselves are valuable artifacts that need to be preserved and, frequently, a data-carrier corresponds to a uniformly catalogued work. Personal Digital Archives of persons of interest to the heritage community often contain manually created hand-held data-carriers. Like the other non-commercial data-carriers, they can vary greatly in structure and quality, but also form valuable artifacts in and of themselves. They typically don't require mass stabilization but may be analyzed in detail for nuances, even for deleted files. These differences result in different approaches to stabilizing their content.

In this paper we explore the experiences gained during our work on the Endangered Archives Programme (EAP), with a particular focus on optical disk processing (rather than tapes, external hard-drives, etc.). All together, the EAP collection currently contains 67 terabytes of data, on approximately 18,000 optical data-carriers, with more to arrive for a further 8 years. In the future, a larger proportion will arrive on more efficient data-carriers, such as external hard-drives. But this was not always practical in the past as early external hard-drives were not found to be robust enough to survive transport and indefinite storage. Much of the material has therefore been submitted via writable CD-ROM and DVD disks, and manually handling these data-carriers has proven to be enormously time consuming. It is, therefore, important to develop scalable processes, and to establish efficient, standardized ways of recording metadata and to deal with defective data-carriers.

The goals of the Endangered Archives Stabilization Project were:

- to move the digital material from hand-held carriers onto backed-up online storage in order to stabilize it,
- to develop workflows so that future accessions can be immediately made available on online storage,
- to determine the best technical set-up, processing approach and software solutions,
- to determine staffing needs for the stabilization process.

We have experimented with different disk-copying robots, developed stabilization processes, defined metadata, and addressed storage issues to scale stabilization to the large quantities of digital objects on hand-held data-carriers that need to be preserved.

## 2 MIGRATION STRATEGY

In general, we wish to maintain the authenticity of the original item as closely as possible. Ideally, therefore, we would aim to perform a reversible migration, such that the digital entity we create from the original data-carrier could be used to create a new data-carrier that is functionally equivalent to the original.

To understand how this might best be achieved, we first summarise how data-carrying media are designed. In order to function, any data-carrier capable of carrying more than one bitstream must use some kind of container format to arrange the bitstreams on the carrier in a way which allows them to be reliably disentangled when read. This is achieved in a range of different ways depending on the media, e.g. by using disk partition maps and file-systems. Necessarily, in order to allow the bitstreams to be distinguished, these container formats must also specify some metadata such as the filename associated with the bitstream, where on the disk it can be found, and how big it is. Usually, the container metadata also includes checksums and error-correction codes to help compensate for any bit loss during creation, aging or use of the media.

By definition, it is impossible to extract the individual bitstreams from the carrier without also stripping away the container. If we are fully confident that we are aware of all of the potential metadata that we might wish to keep, then this information can be extracted along with the bitstreams. But evaluating the auxiliary container-level metadata is time consuming, and if we are forced to make this evaluation directly from the physical media then the media handling process becomes extremely difficult to scale.

Fortunately, this bottleneck can usually be avoided by creating disk images. Here, rather than extracting bitstreams directly to files in a new file system, we attempt to extract a single large bitstream that represents a precise copy of both the contained bitstreams and the container. In this way, we preserve the logical content as completely and as closely to the original as possible. Note this does not necessarily preserve the precise physical layout of the data. For example, a hard disk cloned in this manner will contain the same information as the original, but will not have the same degree of data fragmentation, as the block-level data layout will have changed. However, the clone is logically equivalent to the source disk, and this is entirely sufficient for our purposes.

Critically, this approach also allows us to proceed quickly, migrating the content as soon as possible while minimizing the risk of discarding any data or metadata during bitstream extraction or container transformation. By creating a disk image we can move the original submissions onto safer storage without compromising the authenticity of the originals. This approach is also common in the digital forensics area, and well-established practices are in place for many types of media [7].

### 2.1 Variations in Carrier Type
While the broad strategy of making disk images is a sound one, there are a number of practical difficulties implementing this ap-proach due to the variations in the types of disk and the degree to which the disks conform to the appropriate standards.

The variation in disk formats arises due to the complex history of the medium, and the ways in which the form has been extended or modified to cover different use cases. The original Red Book [12] standard from Phillips specified how to construct a digital audio compact disk, with raw audio bitstreams arranged into a series of session and tracks, along with the physical layout and analogue tolerances to which this format should be constructed. In the following years, a wide range of other standards were published (the so-called Rainbow Books [30]), covering extensions and modifications to this base format, such as CD-ROMs for data, mixed-mode audio and data disks, extended embedded metadata, technical protection mechanisms, and so on.

Since then, and in reaction to the complexity of this group of standards, the vendor community has worked to standardize the way in which the data is laid out upon the disk, via the Universal Disk Format [16]. Both DVD and Blue Ray media use this disk format, which specifies just one container format, but captures the different media use cases in the standardization of the bitstreams within the container, rather than via the structure of the container itself. This is not done for reasons of preservation, but for reasons of ease of creation. Working with a single image makes disk mastering much more manageable. However, this convergence is also extremely welcome from a preservation point of view, as a single class of disk image can be used to cover a wide range of media.

For our older material, we must be able to cope with this variation in form, and even for newer materials, we need to be able to cope with the common variations in the way in which the media conform to the standards. This is particularly true for consumer writable media, where the software that creates the disks does not always behave reliably. This manifests itself not just as systematic deviation from the standards due to software or hardware problems, but also as variability in the quality of the disks due to the reliability of the creation process. For example, when creating ('burning') a writable CD, the process can fail and create unreadable disks (known as 'coasters'), particularly when the disk creation speed is high. For this reason, optical disks should be checked immediately after creation, but this is difficult to enforce when working with external parties.

With some assistance from the curators we were able to identify some particularly 'difficult' collections, and used those as a starting point to determine what type of variation there was in the optical media format. Across our collections, we encountered a very wide range of disk formats on optical media:

- DVD [8] or CD-ROM [9, 13] data disks in ISO 9660/UDF format (containing TIF, JPG, audio data files, etc.).
- DVD Video [8] disks in ISO 9660/UDF format [17, 18] (containing video data, e.g. VOB files).
- HFS+ (Mac) [3] format data disks.
- Red Book [14] Audio disks with sessions and tracks
- Yellow Book [13] Mixed-mode compact disks with a leading or trailing ISO 9660 data track containing mixed media alongside the audio tracks.
- Malformed 'audio' disks arranged in audio-like tracks, but the tracks themselves containing WAV files instead of raw CDR data.

The ISO 9660 specification [17] defines a disk image file format that can be used to clone data disks. This approach gives one single archive file that includes all the digital files contained on a CD-ROM, DVD, or other disk (in an uncompressed format) and all the file system metadata, including boot sector, structures, and attributes. This same image can be used to create an equivalent CD-ROM, and indeed mastering data disks is one of the purposes of the file format. It can also be opened using many-widely available software applications such as the 7-Zip file manager [15] or the WinRAR archive shareware [21].

Similarly, for later disks, such as DVDs, the UDF disk format specifies the layout of DVD disks and a general DVD image file format which is backwards-compatible with ISO 9660. The situation is similar for HFS+, as the data can be extracted as a single contiguous disk image without any significant data loss.

While CD-ROM, DVD and HFS+ format disks are reasonably well covered by this approach, there are some important limitations. For example, the optical media formats all support the notion of 'sessions' – consecutive additions of tracks to a disk. This means that a given carrier may contain a 'history' of different versions of the data. By choosing to extract a single disk image, we only expose the final version of the data track, and any earlier versions, sessions or tracks are ignored. For our purposes, these sessions are not significant, but this may not be true elsewhere.

For DVD disks, the main gap is that the format specification permits a copy protection system that depends on data that is difficult to capture in a disk image. Specifically, a data signal in the lead-in area of the disk contains information required to decrypt the content, but most PC DVD drives are unable to read this part of the disk, by design. Fortunately, this does not represent a problem for the Endangered Archives content, as it does not rely on media that use DRM or other technical right restriction methods.

The situation for Red and Yellow Book Compact Disks [13] is significantly more complex. As mentioned above, the overall disk structure, the sessions and tracks that wrap the data, are not covered by the ISO 9660 file format. Furthermore, it does not capture the additional 'subchannel' data that lies alongside the main data channel, which is used for error correction, copy protection and more esoteric purposes (see the CD+G standard [31] for an example). This information is often hidden from the end user, and indeed many CD drives are unable to access subchannel data at all.

Any attempts to preserve the full set of data channel, session and tracks is inhibited by the fact that there is no good, open and mature file format to describe the contents of a CD precisely. Proprietary and ad-hoc formats exist, but none are very widely supported, standardized or even documented. Even for simple Red Book Compact Disk Digital Audio media, there is only one standardized, preservation-friendly format that accurately captures the session, tracks and gaps – the ADL format [26]. This is a relatively new standard, and is not yet widely supported. Given this situation, we were forced to choose whichever file format is the most practical in terms of the data it retains, given the types of content we have, and by how well the tools that support those formats can be integrated with our workflow.

## 2.2 Disk Images Choices

Due to the variation in media formats outlined above, our overall migration workflow must be able to identify the different cases and execute whatever processing and post-processing steps are required. The first decision we must make, therefore, is to decide what type of disk image we should extract for each type of disk.

If the original data-carrier is a valuable artifact, data-carrier disk images should be produced and treated as the preservation copy.

Similarly, if file system metadata contained in a disk image may contain significant characteristics of the digital object that should be preserved (as is the case, for example, for bootable magazine cover disks) then the disk image should be treated as the primary preservation object. Ideally, this should be in a format that captures all of the data on the disks, not just the data from the final session.

In contrast, if the data-carrier could be considered simply a transfer medium and direct access to the data files is desired, they can be extracted as simple files instead. However, as indicated earlier, this can only be done once the data-carrier metadata has been properly evaluated, so practically we extract as full disk images at first, and then carefully generate the preservation master files from that image. Thus, we decided that all CD-ROM and DVD data or video disks should be ripped to ISO 9660/UDF disk images.

Similarly, the HFS+ disks should be ripped as single image bitstreams containing the volume data. These also manifest themselves as disks containing a single data track, but that happens to be HFS+ formatted instead of using ISO 9660/UDF. Therefore, the process of extracting the data track is identical to the previous case, and the difference lies only in the post-processing procedure.

Unfortunately, the Red Book, mixed-mode Yellow Book and malformed disks could not be extracted to ADL, as the available tools did not support that format. Those tools only supported the proprietary Media Description (MDS/MDF) file format (no public specification), which limits the range of post-processing tools we can use, but which can contain all the information on the disk and thus could be migrated to a format like ADL in the future.

For the Red Book disks, the content of the MDS/MDF disk image files can then be extracted in post-processing with extraction software such as IsoBuster [23]. Unlike the other extraction software we experimented with, IsoBuster could identify and read the full range of disk images we encountered, including HFS+ disk images. The breadth of formats supported was the main reason why IsoBuster was our preferred tool for post-processing MDS/MDF disk images.

Unfortunately, IsoBuster was not able to extract the audio track data reliably when operated in batch mode, and we found the most robust workflow was to use IsoBuster to migrate the disk image to another format with broader tool support. This second image format is known as CUE/BIN format (no public spec), and consists of a pair of files where the cuesheet is a simple text file describing the tracks and their arrangement on the disk, and the binary file contains the concatenated data from each tracks. This format is therefore less comprehensive that MDS/MDF, as the sessions and subchannel data have been discarded, but allows other software such as bchunk [11] to be used to produce usable WAV files from the raw binary data. The mixed-mode Yellow Book disks ripped in the same way as the audio disks, but extracting the content is slightly convoluted. After using bchunk to extract any ISO 9660 data tracks, each must be further processed to extract the files.

The malformed disks can also be ripped to MDS/MDF format, but complicate the content processing workflow further. After bchunk has been used to extract the tracks, they must be characterized using the 'file' identification tool [5] to see if they contain the RIFF header indicative of a WAV bitstream.

## 2.3 The Robot and the Automation Stack

The fundamental limitation on the throughput of the migration process is the manual handling process; the moving and cataloging of disks, and the opening and closing of jewel cases. Critically, the EAP disks are individually labeled by hand, were kept in sets associated with a particular project, and the ordering of the disks had to be retained. When processing the disks, the association between the physical item and the electronic image must be maintained, and so the overall workflow must ensure that the disk identifier is captured accurately and can be associated with the right disk image. The design of the media processing workflow took these factors into account and optimized to usage of the available staff effort while minimizing the risk of displacing or exchanging any disks.

Originally, we started working with a very large-scale disk robot: an NSM 7000 Jukebox (see e.g. [6]) fitted with 510 disk trays and 7 drives. While in principle such a large machine should allow high-throughput migration of optical media, there were a number of issues that made this approach unsuitable. Firstly, while the hardware was essentially sound, the accompanying software was intended for writing to a pool of disks, rather than reading a stream of disks. Trying to make the machine run 'in reverse' was extremely cumbersome, and such attempts were rapidly reduced to firing SCSI commands directly to the disk robot and ignoring the supplied software stack almost entirely. The details of the hardware design also worked against us. For example, the cartridges and disk trays used to load the machine had been optimized for storing sets of disks on shelves in the cartridge after the data had been written to them. This lead to a very compact physical design, but made the process of loading, unloading and reloading the cartridge with fresh disks rather awkward and error prone. In fact, all the robot solutions we looked at were primarily designed for the mass-write use case, but the NSM 7000 support for large-scale reading was particularly lacking.

Putting the media loading issue aside, we found that the main efficiency problem arose from the way exceptions, i.e. damaged, malformed or unusual disks were handled. If all the disks were perfect then the large-scale solution could be made to work fairly easily, with the operator loading up the machine and then leaving it to process a large number of disks autonomously and asynchronously (during which time the operator could perform other tasks). However, a small but significant percentage of the media we have seen have some sort of problem, and so the efficiency of the overall workflow is critically dependent upon this exception rate. This is because the task of reliably picking the problematic disks out of the whole batch rapidly becomes very difficult and error prone when the batches are large. It was, of course, of importance not to misplace or exchange any of the disks from the original collection, and a complex exception-handling process makes this difficult to ensure.

The British Library Sound & Vision group, in comparison, has successfully processed large amounts of compact disks using a single workstation with an array of 10 disk drives. This works well because, as all items are clearly distinct, individually catalogued and barcoded, they can be scanned and processed rapidly. As each disk represents an independent work, the fact that a manually loaded drive array will tend to process disks asynchronously (and thus not retain the disk order) is not a problem. Any problematic disks only occupy a single drive, and the others can continue to be loaded without blocking. Unfortunately, this approach was not well suited to our content, due to the order of batches of disks involved, and the manual cataloguing required per disk.

Following these experiences, we moved to using much smaller robots, the DupliQ DQ-5610 [1] and the Nimbie NB11 [2]. These small-scale machines are only capable of processing tens of disks, but by breaking the collections up into batches of manageable size, the exceptions can be handled more gracefully. This size limitation can then be overcome by running more than one machine in parallel, allowing the process to be scaled up quite effectively as each batch is processed independently. Any exceptions encountered can be tracked more easily, and brought together into a single, manually inspected batch.

Both units are USB 2.0 devices comprised of a single CD/DVD drive and a robotic component that handles the disks. However, the precise physical mechanism is different. For the DupliQ, a robotic arm grips and lifts the disks by the central hole, passing them from a lower tray to the drive and, once the disk has been processed, from the drive to an upper tray, giving a Last-In-First-Out (LIFO) processing order. The Nimbie has a simpler mechanism, with the disks held directly over the drive tray and released by a turning screw, leading to a First-In-First-Out (FIFO) processing order. In general, we found the Nimbie mechanism to be more reliable, as the DupliQ gripper mechanism would frequently fail to grip disks, could not cope with disks sticking together, would sometimes drop or even throw disks, and following this type of hardware error, the software would usually cope poorly and the batch would have to be restarted. Also, the DupliQ can only be loaded with about 25 disks, whereas the Nimbie can be loaded with up to 100 disks, and due to the FIFO ordering, can be run continuously if necessary.

Both small robots also came with appropriate software that supported extracting the disk data as disk images, called QQBoxxPro3. Unfortunately, this proprietary software also forced us to adopt the MDS/MDF format, as this is the only format it supports for multi-track/session disks. However, a more significant limitation was the lack of configurability for different types of disks, meaning that we could not instruct the robot to rip the contents of the different types of disks in different ways, as we would ideally like.

The DupliQ robot came with version 3.1.1.4 of the QQBoxxPro3 software, which appeared to assume that any single-track disks should be ripped as ISO 9660/UDF data disks (with an '.iso' file extension), whereas multi-track disks were ripped as MDS/MDF disk images. This was helpful for data disks and DVD Videos, but potentially quite dangerous for the HFS+ format disks, as that format is quite similar to the ISO 9660 format, and so some software tools open up the disk image and attempt to interpret it as ISO 9660 data without any warning. This makes is appear as if that data is corrupted and/or missing.

The Nimbie robot came with a later version of QQBoxxPro3 (3.3.0.5), which instead simply ripped all disks as MDS/MDF images. This leads to more complete disk images, but means that all of them require significant post-processing to access the data.

For our use pattern it turned out that using the DupliQ's version of the software with the Nimbie robot created the most effective configuration.

It is worth noting that the choice of disk-copying robot can depend on the composition of disk formats in the collection that were discussed in Section 2.1. We ran samples of difficult files on another FIFO robot, the MF Digital Data Grabber Ripstation [10], whose hardware is very similar to the Nimbie. The software on the 2 robots produced useful images for different disk carrier variants, ejecting disks in different situations and left differently useful log information that permitted identifying the presence of problematic situations. Depending on the expected distribution of disk file formats, one or the other of the two robots would have been preferable.

## 2.4 Disk Copying Workflow

Batches of disks were received from the curatorial group, and placed in a safe location next to the disk processing station. This station consisted of a basic PC with the USB robot, and a number of other items listed in Section 5 below. Large sets of disks were broken down into manageable batches of around 30 disks. Initially, this was because of the batch size restriction of the DupliQ machine, but due to the manual-handling overhead introduced by the problematic disks, this relatively small batch size was retained so that the exceptions could be managed effectively.

For sets of disks with low exception rates, the FIFO nature of the Nimbie machine proved very useful, as larger sets of disks could be continuously loaded into the machine. Of course, having multiple machines allows the processing of separate batches to be parallelized, and we found this to be a very effective approach. Initially, we set up two processing stations, running a DupliQ unit and a Nimbie in parallel. However, working with the two different disk-processing orders (LIFO v. FIFO) meant performing two different cataloging processes, one noting the disk identifiers in reverse, recording the disk metadata became a risky procedure. Furthermore, as indicated above, the DupliQ hardware was slightly less robust and reacted badly to difficult disks. Therefore, we moved to running two Nimbie units in parallel instead.

With these two processing stations running in parallel we were able to achieve processing rates of 1,050 disks per month, corresponding to data rates of 2.2 TB per month. The parallel robots had allowed us to minimize the fraction of the time spent waiting for the processing to finish, before the next load of disks could be processed, while overlapping the manual handling with the extraction process as much as possible. The limiting factor in the process at that point was the need to manually create metadata.

## 2.5 Handling Defective Disks and Other Exceptions

We found a range of particularly problematic disks, with the majority of them being physically malformed in some way that made them unreadable. In some cases, this manifested itself as disks that hung for long times in the drive, after which they were reported as being unrecognized. In others, the extraction would start as normal, but would slow down and eventually hang due to some local disk error. In rare cases, manual recovery of these disks was possible just using a different combination of drive hardware and ripping software. Usually, however, our only option was to make the curators aware of the issue as soon as possible, so that they could get in touch with the original authors promptly and get the content re-submitted on new media. In general, we found that disk problems or failures were correlated, i.e. most projects would have no problems, but some would have many problems. It was not possible to determine the root cause of these problems, but clearly systematic failures during the disk-burning process seems to be a more likely cause than simple disk aging or bad disk batches due to manufacturing defects.

Sometimes the cause for stabilization failures was unrelated to production properties of the disk or files. Problems with statically charged disks could be handled by attaching anti-static straps. Similarly, dirty disks tend to stick together. Since the EAP disks are not valuable artifacts that need to be protected in the long run and since our disk drives were inexpensive and did not justify thorough disk cleaning in order to protect the drives, we did not rigorously clean disks as a matter of principle. If visual inspection showed dust we used a powerful camera lens cleaner to blow it off. A compressor proved to be too noisy for the shared office environment. More stubborn dirt was washed off using a solution of distilled water and isopropyl alcohol in equal parts. We used camera lens microfiber cloths for wet and subsequent dry cleaning and were careful to clean disks radially from the center of the disk straight to the outer edge in order to avoid inadvertently scratching consecutive data. We received valuable advice from the British Library Sound and Vision studios on disk cleaning issues. In one case, the cause of stabilization failure turned out to be labels that were affixed to the disks. The paper used for the labels was too thick and caught in the disk drive. We placed moist cloth onto the labels in order to soften and peel them off. For severely scratched disks we were able to borrow a disk polishing machine that physically polishes off some of the disk's thickness to remove scratches.

## 3 CONTENT MANAGMENT

Stabilization is only the first step in improving the preservation quality of the content submitted on hand-held data-carriers (see Table 1). Once the material is stabilized it experiences changes for curatorial, preservation or access reasons. This includes identification of duplicate, damaged or problematic images, securing replacement images, re-organization of content to standardize the collection structure, cataloguing, generating access surrogates, and the embodiment of these processes and relationships as archival metadata. It is striking that this process can extend over a long period of time (possibly years) due to the large volume of projects, projects continuing to submit materials, and the manual intensiveness of the curation.

Figure 1 illustrates the content management evolution that will be discussed in the following sections.

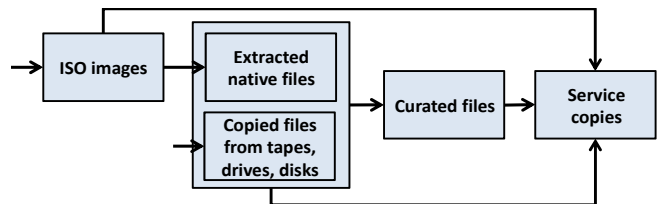

**Figure 1: Overall content evolution, blending resources from different carriers into a curated, accessible artifact**

This multi-stage workflow creates multiple versions of the same item, and the pressure this creates on the available storage space means that that previous versions need to be overwritten in order to keep costs down. This means that the process needs to be extremely well documented, so that ancestor items can be purged with confidence. Furthermore, the long timeframe of each project means that we must be able to add content over time, and our data management procedure must be able to cope with this. We must, therefore, diligently collect a range of metadata during data processing in order to ensure the content can be managed well over time.

For example, a fully stabilized data-carrier should consist of a set of files (disk image or content files) with associated metadata, linking the online content to the physical data-carrier via its identifier, linking to other metadata provided with the content, and documenting the stabilization event as part of the content's provenance. We chose to record all metadata in simple Excel spreadsheets that enforce controlled vocabularies and detect duplicate values when appropriate. When designing the spreadsheets we made sure that all fields could later on easily be translated into the final METS [28] and PREMIS [29] formats we tend to use. Further details of the metadata we chose to capture can be found in section 4.

### 3.1 Content Evolution

The overall content workflow shown in **Figure 1** is built up from a chain of more fine-grained processes and actions. In this section, we discuss some of those processes and the particular preservation issues that arose during their implementation.

*3.1.1 Selecting the Data-Carriers to be Stabilized*

Curators select the data-carriers that should be stabilized next based on prioritizing and balancing the following factors:

- risk assessment (e.g. age of the data-carriers, expected level of problems with reading the data-carrier, or expected difficulty of re-requesting material for faulty data-carriers),
- quick gain (process large external hard-drives over DVDs over even smaller CDs),
- user demand for the content.

If there are duplicate data-carriers (possibly at different resolutions) the curators should chose the highest quality copies and, if this is not obvious, ask digital preservation advice. Simple rules of thumb, such as preferring a TIFF file to a PDF, don't always apply. For example, we found low-resolution files with duplicate PDF's that contained higher-resolution TIFF duplicates.

*3.1.2 Stabilize the Data-Carriers*

The content migration itself, as described in Section 2, is embedded in a physical media management workflow. The disks chosen for stabilization are transferred to the stabilization station and the location of the data-carriers is recorded on a dedicated spreadsheet. This is usually a single large batch corresponding to a single project. This batch is then broken down into smaller batches, so that they can be processed efficiently. Physical place markers are used to keep track of each batches location in the relevant storage boxes. Processed disks are returned to this location, whereas failed disks are replaced with colored disk sleeves that contain a written record that identifies the disk, the nature of the failure and what has and is being done in order to resolve the failure. Stabilization metadata (as described in Section 4) is created as the robots are being loaded. Unidentified or inappropriately identified disks receive a unique data-carrier identifier which is recorded in the disk hub and the metadata spreadsheet. At this step, we also physically clean disks, should this be necessary.

*3.1.3 Clean up Disk Image Directories*

Clean-up of directories containing disk images can occur once stabilized batches of disk images are combined into one online project. It involves removing metadata, directory structures and files that were previously only needed to manage the stabilization process and merging and de-duplication of the batched files. We have decided to execute the latter process in a just-in-time fashion, i.e. when the following extraction step is to be executed, rather than every time a new batch is added.

At this point we can also execute the post-processing of disk images for the rarer data-carrier variants that was discussed in Section 2.2. If the disk-copying robots by default produced output formats that were not optimal for the data-carrier variant, this is recognized at this point and the actually desired output disk image formats are created. For example, we create non-proprietary WAV/ADL files from proprietary MDS/MDF files. Again we need to create checksums for the newly created files.

### 3.1.4 Extract Content for Curation

In order to permit content curation to take place, an additional goal was to make data files individually and uniformly accessible. Disk images, tape and hard-disk representations received from the original suppliers are represented differently and require different modes of access. While the desire to preserve as authentically as possible drives us to store disk images, this form should be made transparent to the user so that they can quickly and reliably access the files themselves. Representing files universally in a uniform file-system and directory structure with uniform structural metadata makes it possible to access them all in the same way.

For disks we can use software tools such as IsoBuster [23] to extract the files from the disk images. The stabilization of digital objects from tapes or external hard drives best happens directly as files in their native file format since image formats and image extraction software for them are not in wide spread use (particularly for the MDS/MDF format, as noted earlier).

Where disk image formats are available, alternatively, data files can transparently be extracted on demand using software appropriate for the disk image.

### 3.1.5 Select and Create Curated Folders and Files

In the cases where data-carriers are valuable artifacts in their own right, the disk image is considered the preservation copy of archival value. No further curation of the content is desired.

In the cases, however, where the data-carrier mostly functioned as a means of data transfer, curators may change the content files. They may reorganize folder structures, move files into a more logical order helpful to human interpretation, rename files for better human consumption and link to those files from their catalogues. They may discard any unwanted directory structures, delete or replace unwanted files, duplicate files or files that don't satisfy collection guidelines. This is a manual process, and these curated files and folders have enormous value added compared to the files merely extracted from the data-carriers, and therefore must be considered the ultimate preservation copy.

In addition, individual files may be changed for preservation reasons. One may migrate files that are not in a preservation-worthy format (e.g. proprietary raw files that are not supported, password protected files, etc.) If this is done, then the original item is also kept unless we have a very high degree of confidence that the migration was as sufficiently complete.

### 3.1.6 Generate Service Copies

Once files in their native file formats are available one can derive service copies for user access. Curators select the files from which service copies should be created for presentation to the end-user and staff responsible for the service copies generate them from the preservation copies. While they should be backed up and managed with customary care, they are not part of the digital preservation cycle. They can be recreated from preservation copies if the need should arise.

## 3.2 Storage Issues

The large size and complexity of the content and its evolutionary nature led to several network storage issues.

Most importantly, the disk images had to be stored alongside the content extracted from hard-drive submissions, and the curated versions of the content for each project. Additionally, some projects submitted material on mixed data-carrier types. We managed the relationships between the collection parts that are stored as disk images and those that are stored in the native file system in two ways. We recorded their structural relationships in metadata and we used uniform directory naming conventions that indicate how the parts of the collection relate. For example, we may have a collection of files stored in the file system that is derived from a collection of disk images. This derivative relationship is recorded in the project status spreadsheet and also expressed through directory naming conventions. Its directory structure looks different from the situation where half of the collection is submitted on disks that were stabilized as disk images and half of the collection was submitted on external hard-drives that were extracted as files, therefore forming a sibling relationship. This required clear folder naming conventions to be decided and enforced.

In practice, cleaned up versions or curated folders can over-write earlier representations of the same content. For now, we have a cooling off period for each collection in order to determine whether any problems are likely to arise. After this digital preservation experts and curators together decide whether older representations should be deleted.

Content that evolves slowly over time is a challenge in the presence of a write-once digital repository. In this case, a reliable intermediate storage architecture needs to be determined.

## 3.3 File Management Issues

In addition to the content management steps described above one needs to perform a large number of common file management tasks. These include

- creating checksums (discussed in the following sub-section),
- copying large batches of files across the network,
- merging and de-duplicating batches of files,
- managing and keeping track of the various batch locations that were created due to limitations in the maximum size of an available storage unit (about 8TB),
- renaming and moving files and directories whilst recording the relationship to the old names and folder relationships,
- identifying file formats,
- validating files and discovering corrupt files.

The biggest problem we found is that there does not seem to exist a risk-proof tool-kit of stable, uniformly applicable tools that can be handled by non-technical staff. Ideally many of the file management activities should be in the hands of the curators rather than in those of IT personnel or digital preservation staff. But having to customize scripts, having to choose different software tools for quite similar situations, and not being able to mask "dangerous" flags in operating system commands currently still require specialist involvement.

As an example of this issue, the following sub-section discusses creation and management of checksums.

### 3.3.1 Create Checksum Manifests

All disk images (and indeed all content) should have a verifiable manifest of checksums created as soon as possible during the item's lifecycle. This can then be used to ensure that no items have become corrupted or lost over time. Therefore, as soon as possible after the disk images were written to the external hard-drive, a full manifest of checksums was created for the complete hard-drive. The disk images were then transferred to backed-up network storage, using the checksum manifest to verify the transfers. Note that we chose to use SHA-256 checksums [21], as these are the type used by the institution's archival store and so can accompany a digital item throughout its entire lifecycle.

One problem we encountered is that checksums produced by different content creators or by different software can contain different variants (such as checksums run in binary or text mode) but that this is not necessarily indicated correctly in the manifest. Similarly, different tools use different text encodings and/or folder separators in paths (i.e. forward slashes or Windows-style backslashes). Additionally, the checksums can be laid out in different ways (e.g. per file, folder or collection). We choose top-level collection manifests so that completeness can be managed along with integrity, but this mode of operation was not well supported by all tools.

Another, more serious problem was these tools were not sufficiently robust or user-friendly. A range of GUI programs do exist, but most were found to have some circumstances when they did not behave as expected, or were difficult to use. This is of critical importance when we wish the curators or content creators to build the manifest early in the lifecycle. The best tool we have found so far is the ACE Audit Manager Local Web Start Client [24]. This can be launched easily and has a reasonable user interface. Ideally, this tool might be extended to allow more formal data transfer and storage structures (e.g. bags as per the BagIt specification [19]) to be generated from sets of files with known hash sums.

## 4 METADATA

We managed three types of metadata pertaining to the data-carriers we stabilized. The primary and secondary metadata described below are of a descriptive nature and are of particular importance for supporting the curatorial work. The stabilization metadata is the metadata we produced in order to support the stabilization process.

### 4.1 Primary Metadata

Primary metadata is provided by the projects to describe the content and to describe the data-carriers and their structure.

#### 4.1.1 Submitted Metadata Listings

Metadata listings are submitted by each project separately from the content data. Most of them are submitted as one or more separate Microsoft Word or Excel documents, sometimes in the form of an Access database or in the form of a paper document. The metadata listings are recorded upon receipt and linked to the data-carriers via the project and data-carrier identifiers.

There is also non-digital metadata such as handwritten notes inserted with or written on data-carriers.

#### 4.1.2 Target Metadata

Curators catalogue all projects using the ISAD(G) and ISAAR(CPF) descriptive metadata standards down to archival file level, sometimes item level.

### 4.2 Secondary Metadata

Besides explicit listings and content descriptions that are identified by the projects as primary metadata, there are also additional PDF, Microsoft Word documents or JPG images mixed in with the digitized content that document the circumstances, location or people associated with the collection. There is also some descriptive metadata included in the set of content data, e.g. a handwritten note giving descriptive or manifest information, which has been scanned and, together with the content, included in the submission. We even found a file that contained the password for accessing the remaining files in the folder.

It is currently not possible to detect this born-digital secondary metadata automatically from within the content files; rather it can only be identified manually since it requires curatorial judgment.

### 4.3 Stabilization Metadata

Stabilization metadata are the metadata produced as a result of the stabilization process. Stabilization metadata are created in order

- to document the provenance of the content files to provide information on the content's authenticity and to enable problem solving if a tool involved in the stabilization process should have caused problems,
- to link the resulting content files to their original data-carriers: Disk images or extracted files should be linked to the original data-carrier so that it can be found if there are problems with the extracted content or if questions need to be directed to the project that has submitted the material,
- to link the content to its primary and secondary metadata.

Additionally it must, at later processing stages, record:

- if directory or file names are renamed in order to be able to link to previous versions,
- if the bit-representations of files or images are changed to document the degree of authenticity of the content,
- from which preservation copy, and how, a service copy is derived to document the degree of authenticity of the delivered content.

We documented the following types of stabilization metadata:

**Data-carrier identifier.** Data-carriers may be uniquely identified through such information as project number, accession numbers, box numbers, disk order within the storage box, unique data-carrier identifier within the project that were assigned by the project, cataloguing codes, etc. We set up the metadata spreadsheets to flag up duplicate data-carrier identifiers, recorded duplicate naming events and created disambiguating identifiers. Data-carrier identifiers enable us to link disk images or extracted files to the original data-carrier.

**Metadata about where the data-carriers are located during the stabilization process and who is responsible for them.** They may be in their regular shelf space, at the stabilization processing workstation, separated out for further investigation of failures, or with a curator.

**Metadata about the stabilization event.** This is provenance metadata that keep track of what processes the content has undergone. We record for each project which data-carrier has been stabilized and to what degree of success:

- When the data-carrier was stabilized,
- Who stabilized the data-carrier,
- What software and hardware was used to stabilize the data-carrier,
- Status of the stabilization event. This may be closed-Successful, closed-Manual clone, closed-Partial clone, closed-Failed, open-Partial clone, open-Failed, Not attempted,
- Primary metadata accompanying the data-carrier. This
    o may be transcribed into the metadata spreadsheet,
    o may be an image scan or photo of the accompanying documentation,
    o may be linked to a primary metadata register via the unique data-carrier identifier,

- File names of the output files of the stabilization process, i.e. the names of individual disk images,
- File extensions of the output files of the stabilization process,
- Administrative metadata to help manage the handling of the batch, such as the number of disks per stabilization batch, how many attempts have been made, run time of the stabilization, comments that explain special circumstances, etc..

**Metadata recording the project status.** This records which stages of the processing workflow the project has undergone (when and using which software and hardware) following stabilization. This includes the processes outlined in section 3.1.

**Storage information.** This records location and quantity of the stabilized content.

# 5 APPENDIX: STABILIZATION WORKSTATION REQUIREMENTS

Setting up a fully equipped stabilization workstation may include the following items. In each case we list the item, why it is needed; and the instances we used without any intention of endorsing the particular product.

## 5.1 Hardware Components

**PC**: For cloning disks with disk copying robots onto external hard-drives, editing metadata, check-summing; Dell and HP standard issue PCs with Microsoft Windows XP.

**Disk-cloning robot**: To automate disk-handling, Nimbie NB11 disk robot.

**External hard-drives:** To store cloned images before transfer to the server. As large as possible, formatted as required by the wider environment (i.e. NTFS for our Windows environment).

**Server with large storage:** For copying digital objects from external hard-drives, processing stabilized digital objects and for intermediate storage; Microsoft Windows Server 2003 R2, Standard Edition, Service Pack 2, Dual Core AMD Opteron.

**Mass storage:** For medium-term storage of digital objects before ingest into the permanent digital repository.

## 5.2 Software Components

**Spreadsheet software**: For capturing metadata; MS Excel.

**Cloning software for robot:** To clone disk images; QQBoxxPro3 v. 3.1.1.4.

**Cloning software and software for extracting files from ISO images (by hand):** For quality assurance of cloned image disks and for manual cloning of failed disks; IsoBusterPro 2.8.0.0, (7-zip, WinRar as back-up options).

**Media player software**: To check if the video disk images or the disks failed by the robot can be played; Real Media Player, VLC Media Player 1.1.4.

**Check-summing software**: For creating check sums in order to validate the integrity of cloned images or files when they are stored or transferred; FastSum 1.7.0.452[20], ACE Audit Manager (local WebStart client)[24].

**File transfer software:** For transferring and combining large batches of files; RichCopy 4.0, Windows Explorer, rsync.

**Audit tool**: For periodic checksum validation and for detecting duplicates; ACE Audit Manager (server version) [25].

## 5.3 Additional Workstation Components

**Barcode scanner:** To scan barcodes of already catalogued items in order to link to existing metadata. Not used for EAP.

**High quality disk reader:** To investigate failed disks; Plextor Blu-ray Disk Drive (PX-B120U) and standard PC disk drive.

**Camera or scanner:** For taking pictures of inserted or written information that accompanies the data-carriers. These images form part of the data-carrier metadata.

**Tripod or camera stand:** For holding the camera in place.

**Lighting**: For camera and for inspecting physical disk damage to disks.

**Dust blower and/or oil free airbrushing compressor with nozzle:** For cleaning disks; 1 Giottos GTAA1900 Rocket Air Blower, AB-AS18 Mini Piston type on-demand compressor for airbrushing, compressor hose, Sealey SA334 - Air Blow Gun, airbrush hose adaptor - 1/4"bsp female to 1/8"bsp male.

**Workbench mat:** For catching dust, to prevent it spreading in the office.

**Microfiber cloths, wet and dry, liquid dispenser with 50% isopropyl alcohol**, **50% distilled water:** For cleaning heavily soiled disks; Visible Dust Ultra Micro Fibre Cleaning Cloth.

**Place Markers** For marking batch start and end and problem areas within disk storage boxes; cut from plastic backing of folders, bookmarks.

**Disk sleeves** To mark places in disk storage boxes where failed disks have been removed; into them we insert labels describing reason for, date of removal, etc.; Compucessory CD Sleeve Envelopes Paper with Window.

**CD/DVD marker, xylene free, with, ideally, water-based ink**: To write data-carrier identifiers in disk hub; Staedtler Lumocolor CD/DVD Marker Pens Line 0.4mm Black 310 CDS-9 (which is alcohol based)

# 6 CONCLUSION

We have developed a robust workflow for stabilizing hand-held data-carriers onto online storage. For disks, we considered different approaches of parallelizing the copying process: a large-scale and 3 small-scale LIFO and FIFO disk-copying robots as well as asynchronous stacks of disk drives. We found that a small-scale FIFO disk-copying robot was the best tool for the given collection. We managed to set up a workflow that parallelized the copying to a point where the time needed for copying was balanced with the time needed for the manual tasks of disk handling and metadata creation, thus optimizing our use of the available staff member's time. We also adjusted the process batches so that problematic disks could be handled successfully without upsetting the smooth running of the workflow.

Whether disk images, content files in their native file formats or curated files are to be considered the ultimate preservation copy depends on the nature of the collection. In particular, it depends on whether the data-carrier itself or the curated folders represent the archival artifact. It also depends on the likelihood of having to go back to originals. It may be appropriate to, for example, keep disk images and curated files if storage considerations permit this.

However, the most important consideration when exploring the potential migration workflows was to let the choice of technical solution be driven by the development of the physical workflow and the nature of the content, rather than to choose the technical solution first and try to force the manual process to work around it. We have adjusted the workflow to be as intuitive as possible and have sought to trim away any unnecessary steps. We documented the process as carefully as possible and validated the documentation by asking a new staff member to execute the workflow just from documentation. This has helped us to make the overall process robust and resilient enough that the curatorial team involved can take over the data-carrier migration process and proceed independently.

Most of the difficulties that arose during the development of this procedure originated from the suitability of the available software. For example, it is unfortunate that the disk-copying robots' software was not able to automatically recognize the data-carrier variants and perform different stabilization activities based on them. This meant that we had to write software to identify problematic situations where the wrong output format was produced and remedy the action in a post-processing step. Similarly, another significant obstacle to developing readily usable stabilization workflows was the lack of robust, intuitive file management software that can be handled by non-technical staff without the risk of inadvertently damaging the collections.

Data-carrier stabilization is a very time-consuming task that should be included in the planning for any project that includes hand-held data-carriers. A total volume of 100 TB requires 5 person years to stabilize at a rate of 20 TB/year. Obviously this number varies greatly with the type of data-carrier and the incidence of failed carriers. Staffing needs and the expenditure for stabilization workstations need to be included in business plans. Staff must be detail-oriented and systematic, but also flexible to respond to previously unencountered situations that require novel technical or pragmatic solutions. We had to process a surprisingly large number of projects before new types of workflow exceptions became rare.

Finally, while not a major concern during this project, other work on data-carrier stabilization must consider any copyright issues. We stabilized a collection for which we had the content owners' consent for copying. For other hand-held data-carrier collections at the British Library the copyright issues are difficult and the effort required to sort out permissions prohibits their stabilization.